# Polarized Raman Spectra and Complex Raman Tensors of Antiferromagnetic Semiconductor CrPS$_4$

Sujin Kim[1], Jinhwan Lee[2], Changgu Lee[2] and Sunmin Ryu[1*]

[1]Department of Chemistry, Pohang University of Science and Technology (POSTECH), 50 Jigokro 127, Pohang, Gyeongbuk 37673, South Korea

[2]School of Mechanical Engineering, Sungkyunkwan University, 2066 Seoburo, Suwon, Gyunggi 16419, South Korea

* E-mail: sunryu@postech.ac.kr

## ABSTRACT

Chromium thiophosphate (CrPS$_4$), a monoclinic crystal of $C_2^3$ space group, is a ternary layered semiconductor with an optical bandgap of 1.4 eV and exhibits antiferromagnetism below 36 K. Despite its potential in optoelectronic and magnetic applications, the symmetry of its lattice vibrations has not been systematically studied. In this work, we performed a polarized Raman spectroscopy of bulk CrPS$_4$ using three different excitation wavelengths of 457, 514, and 633 nm. High-quality crystals grown by the chemical vapor transport method were mechanically exfoliated or polished to expose three orthogonal crystallographic facets. Polarized Raman spectra were obtained in parallel and cross configurations by rotating samples about the surface normal to each of the facets. Among 33 Raman active modes (16 A and 17 B) at the Brillouin zone center, 19 (8) peaks observed in the parallel (cross) configuration were assigned as A (B) modes. Complex-valued Raman tensors were determined for 7 major A modes using the angle-dependent Raman spectra of the three facets. The results can also be used in determining the crystallographic orientations of $CrPS_4$ unequivocally.

Keywords: $CrPS_4$, polarized Raman spectroscopy, Raman tensor

**Introduction**

Since the first mechanical isolation of graphene in 2004,[1] various two-dimensional (2D) crystals have received tremendous attention because of their novel material properties that are distinct from those of their bulk forms. Whereas the first several years were mostly devoted to studying metallic graphene in various scientific disciplines,[2-4] the past decade has seen substantial interest and research activities in 2D semiconductors represented by binary transition metal chalcogenides (BTMCs). In particular, $MoS_2$ and other transition metal dichalcogenides exhibited indirect-direct bandgap transition[5] and valley-specific electronic transition in the 2D thickness limit,[6] which led to the discovery of rich photophysical dynamics and proposition of various optoelectronic devices.[7-10] Despite the significant amount of investigations, however, it has been realized that the 2D crystals that have been under scientific scrutiny are only a small subset of the whole 2D material family. Ataca et al. predicted 39 stable 2D transition metal dichalcogenides.[11] More recently, more than 1,000 2D solids were identified among 50,000 inorganic crystals using a massive data mining algorithm.[12] Such theoretical predictions suggest that one can discover other 2D crystals with even more unique properties and functions that lack in the known materials.

Very recently, the first 2D ferromagnetic (FM) behavior has been observed for $Cr_2Ge_2Te_6$,[13] and $CrI_3$.[14] Exhibiting exotic magnetism and unconventional superconductivity,[15] low dimensional antiferromagnetic (AFM) materials are also of great importance and require extensive research. Moreover, the heterostructures of 2D ferromagnets and antiferromagnets cannot only be crucial in resolving the controversy[16] regarding the mechanism of the exchange bias effect,[17] but may also lead to novel findings and applications. Ternary transition metal chalcogenides (TTMCs) are promising candidates for 2D magnetism because of their diverse compositional and structural possibilities. Several metal phosphorus sulfides ($MPS_4$ or $MPS_3$) have recently gained attention again[18-26] decades after their first discovery.[27] While some tetrasulfides (Ga and Bi) and trisulfides (Sn, Zn, and Cd) are diamagnetic, one tetrasulfide (Cr) and other trisulfides (Mn, Fe, Co, and Ni) are paramagnets that become AFM below their Neel temperatures.[23]

Chromium thiophosphate ($CrPS_4$) is a layered material and cleaves well into 2D forms of single (1L) and few-layer thickness.[22] Its bulk form exhibits anisotropic antiferromagnetism with a Neel temperature of 36 K[28, 29] and holds a substantial potential regarding 2D magnetism. A recent theoretical study showed that an FM state is nearly degenerate with the experimentally observed AFM state for the bulk.[30] It was also predicted that 1L $CrPS_4$ is an Ising-type 2D FM material unlike its bulk form,[30, 31] and phase transition between AFM and FM states can be induced for 1L by applying lattice strain.[32] Despite its potential in 2D magnetism and various future applications, many aspects of $CrPS_4$ have yet to be revealed. Since its first synthesis in 1977,[33] there were only a few early studies that observed its AFM behavior[28] and electronic transitions[34] involving core and valence bands. After a long silence over the past years, this material just started to gain attention from both experimentalists[22, 23, 29, 35-40] and theoreticians[30-32] awaiting more extensive investigation. In this regard, a complete understanding of lattice vibrations is imperative for this emerging material, although there have been a few recent experimental studies that reported its Raman spectra.[22, 26, 36, 38]

In this work, we anatomized facet- and wavelength-dependent Raman scattering behavior of bulk $CrPS_4$ using polarized Raman spectroscopy at three excitation wavelengths of 457, 514, and 633 nm. Each of 27 Raman peaks was assigned to one or a combination of 17 A and 19 B zone-center modes belonging to the $C_2$ point group. The Raman tensors of several major modes were determined by extensive angle-resolved measurements for samples with three orthogonal crystalline facets. Distinctive polarization dependence of each Raman peak was exploited to establish a rapid and reliable method to determine crystallographic orientations without full-angle measurements. The lattice vibrations of crystalline CrPS$_4$ reported in this work will be valuable in characterizing its structural details and crystallographic orientation in various forms and dimensions.

**Results and discussion**

**Crystalline structure and principal edge planes.** Individual single layers of CrPS$_4$ are stacked with a weak van der Waals (vdW) interaction,[31] as shown in Fig. 1a. Its bulk crystal belonging to the space group $C_2^3$ in a monoclinic lattice system is spanned by the unit cell (Z = 4) with lattice constants of ***a*** = 10.871, ***b*** = 7.254, ***c*** = 6.140 Å, and ***β*** = 91.88°.[33] Each of the chromium and phosphorus atoms is bonded to 6 and 4 sulfur atoms arranged in a slightly distorted octahedron and tetrahedron, respectively. All sulfur atoms, each connected to one phosphorus, and 1 or 2 chromium atoms, are located on either surface of 1L. As reported in a recent study,[22] bulk crystals cleave well into flat-surfaced thinner crystals, and the ease of mechanical exfoliation into single and few-layers was similar to that of graphite and BTMCs. This fact is consistent with the recent theoretical prediction on the interlayer cohesion energy of CrPS$_4$.[31] As described below, the cleaved surfaces were parallel to the ***ab*** plane shown in Fig. 1b[22] and are referred to as a cleavage plane, rather than a basal plane that is defined to be perpendicular to the principal ***b*** axis. The shortest distance between sulfur atoms in neighboring layers is 3.75 Å, much larger than that of a typical disulfide bond (2.05 Å), which indicates the vdW nature of the interlayer interaction. Because of the low symmetry in its crystalline structure, CrPS$_4$ is expected to show anisotropy in various material properties including inelastic light scattering shown in this work.

Figure 1d shows an optical micrograph of an exfoliated bulk-like CrPS$_4$ sample supported on a SiO$_2$/Si substrate. Its atomic force microscope images revealed that the surface is highly flat and smooth, and the heights of typical steps were multiples of the interlayer spacing (~0.62 nm), as shown in Fig. S1. The thickness of the green and yellow areas (marked with diamond and triangle of magenta color, respectively) was about 120 and 70 nm. The characteristic colors that originate from wavelength and thickness-dependent optical reflection[41] were used to identify areas of target thickness. Notably, the exfoliated $CrPS_4$ samples thicker than several layers exhibited two straight edges that are preferentially aligned with an angle of 67.5° (or 112.5°). As shown by the superposed lattice model based on a recent report,[22] the cleaved edges are parallel to the diagonals of the '*ab* plane' unit cell (Fig. 1b), which allows one to determine the crystallographic a and b axes. The preferential cleavage also implies substantial anisotropy in bonding energy density among various possible edges. To define the direction of light polarization, we assigned X and Y axes to the a and b axes with Z axis being perpendicular to the XY plane or *ab* plane. As shown in Fig. 1e & 1f, we also prepared bulk crystals terminated with the XZ and YZ edge planes for polarized Raman measurements. After mechanical cutting followed by polishing (Fig. S2 and Methods), both of the exposed edge planes showed high level of flatness (Fig. S1). It is notable that the latter gave a much smaller root-mean-square roughness than the former.

**Polarized Raman spectroscopy.** Because of the weak interlayer interactions, the lattice vibrations of bulk $CrPS_4$ can be approximated as those of 1L.[42] Since the base-centered unit cell in Fig. 1a contains four repeating units, one primitive unit cell has $Cr_2P_2S_8$, which leads to 36 phonon modes including 3 acoustic modes. A normal mode analysis revealed that the lattice vibrations at the Brillouin zone center consist of the following irreducible representations that belong to the $C_2$ point group: 17A+19B.[43] All except the three acoustic modes (A+2B) were Raman-active, and each of them could be identified as either A or B mode by the polarization selection rule.[44] The intensity (I) of polarization-selected Raman signals can be written as

$$I \propto |\hat{e}_s \cdot R \cdot \hat{e}_i|^2 \qquad (1)$$

,[45] where $\hat{e}_i$ and $\hat{e}_s$ are the unit vectors for the electric polarization of the incident and scattered light, respectively, and *R* is the Raman tensor of a given mode. Table 1 shows the general forms of the two Raman tensors for A and B modes of $CrPS_4$, the components of which will be determined later. Using Equation 1, one can decide whether the polarized Raman intensity for a given detection geometry will be zero. Table S1 shows the resulting polarization selection rule for the back-scattering geometry, where both of the excitation and scattered beams propagate along a normal to the exposed surface. The first and last letters of a given optical configuration specified in the Porto's notation[46] denote the propagation directions of the incident and scattered beams, respectively. The second and third in the parentheses identify the polarization directions of the incident and scattered photons, respectively. Thus, A (B) modes will be silent when the cleavage plane, XY plane, is excited in the parallel (cross) configuration, where the polarization of the incident and scattered photons is parallel (perpendicular) to each other.

Figure 2 shows unpolarized and polarization-selected Raman spectra of XY-faceted CrPS$_4$ obtained with 633 nm excitation wavelength. All the Raman measurements were performed in a back-scattering geometry with the excitation beam normal to the cleavage plane and $\theta = 0°$, where $\theta$ is the angle between the Y axis and the polarization of the incident beam (blue double-headed arrow in Fig. 1d). The scattered photons were detected with or without an analyzing polarizer. The unpolarized spectrum (black in Fig. 2) showed 27 peaks in the range of 50 ~ 1000 cm$^{-1}$ on top of a weak photoluminescence (PL) background, the origin of which was not identified. Each of the Raman peaks could be observed in either the parallel, -Z(XX)Z in blue, or cross, -Z(XY)Z in red, configuration in Fig. 2 and was assigned to A or B mode according to Table S1. The polarized detection also revealed that the peak at 116 cm$^{-1}$ of the unpolarized spectrum is a doublet of A and B modes (Fig. S3). Each of 19 A and 8 B modes was designated $A_i$ or $B_j$ in the order of increasing frequency. The assignment does not include 6 peaks missing, possibly because of their low intensity. The peak at 520 cm$^{-1}$ in the cross spectrum is from the Si substrate, not from the $A_{11}$ mode that should be silent.

As shown in the polarized spectra for the spectral region of 220 ~ 340 cm$^{-1}$ (Fig. 3a & 3b), the intensity of each Raman peak varied substantially as a function of $\theta$ (see Fig. S4 for full-range spectra). With increasing $\theta$ by rotating the sample, $A_5$ reached a maximum intensity at 90° and returned to a minimum at 180°. In contrast, $A_6$ and $A_7$ exhibited local minima between 0 and 90° (also between 90 and 180°). For $B_4$, $B_5$, and $B_6$, the change in intensity was repeated every 90°. It is also notable that A (B) modes can be seen in the cross (parallel) spectra when $\theta$ is not multiples of 90°, which will be explained below.

To reveal the anisotropic Raman scattering in CrPS$_4$ further and quantify its Raman tensors, we also obtained Raman spectra from the two edge planes perpendicular to the cleavage plane, as shown in Fig. 4. Notably, the XZ plane did not show any of the B modes, unlike the other two planes. Instead, several A modes were observed in its cross spectrum, which is consistent with the polarization selection rule (Table S1). The intensity profiles of the parallel and cross spectra from the XZ plane were also very different from that of the cleavage plane. In the parallel configuration, $A_1$ and $A_5$ that were weak or negligible for the XY plane were intense for the XZ plane, and vice versa for $A_6$ and $A_{11}$. Whereas the parallel spectrum from the YZ plane is similar to that from the XZ plane, its cross spectrum showed only B modes, unlike the XZ plane. Note that the intensity profiles are noticeably different for both edge planes.

Multi-wavelength Raman measurements shed more light on the vibrations of CrPS$_4$. As shown in Fig. 5, Raman spectra of CrPS$_4$ crystals were also obtained with two shorter excitation wavelengths of 514 and 457 nm. Although most of the A & B peaks identified with 633 nm (Fig. 4) were observed, their relative intensities in the polarized spectra varied significantly. Notably, the $A_3$ peak could not be seen at 514 and 457 nm excitation. It is also notable that the line widths of $A_9$, $A_{10}$, $A_{13}$, and $A_{15} \sim A_{19}$ are significantly larger than those of other peaks (also see Fig. 2 and Table S2). Some of these broad features may be an unresolved superposition of multiple peaks, and others may originate from second or higher-order Raman scattering. As can be seen in Table S2, the frequency of $A_9$ matches with twice that of $A_4$, and those of $A_{15}$ and $A_{17}$ correspond to the frequency sums of $A_3$-$A_9$ and $A_6$-$A_8$ pairs, respectively.

**Azimuth-dependence of polarized Raman signals**. To reveal the anisotropic behaviors of the Raman peaks more clearly, we depicted the Raman intensities of $A_1 \sim A_8$ peaks in the polar graphs as a function of the azimuthal angle (θ) in Fig. 6a & 6b. The signals were obtained in the parallel and cross configuration for the XY plane using 633 nm. Black circles are the experimental data, and solid lines correspond to the best fits to the data based on the Raman tensors, as will be explained below. The angle-resolved Raman signals of $A_3$ were very weak and not given. Whereas all the polar graphs for the parallel configuration exhibited two-fold symmetry with two major lobes and additional minor lobes for $A_1$, $A_6$, and $A_7$, all except $A_2$ reached their maximum intensity at the same azimuthal angle. Notably, the signals for the cross configuration had a four-fold rotational symmetry with maximum intensity occurring 45° off from the parallel cases. Interestingly, the B-symmetry peaks gave an identical four-lobed angular dependence but with angular displacement of 45° between the two polarization configurations (Fig. S5).

For the other two crystallographic facets, the azimuth-dependence of the A-type peaks in the parallel configuration exhibited more intriguing patterns, as shown in Fig. 6c & 6d. Whereas all the angular patterns had a two-fold symmetry, $A_2$ and $A_4$ peaks for the XZ plane were distinctive for their four-lobed structures but still with two-fold symmetry. $A_2$, $A_3$, and $A_6$ peaks differed from the rest in their maximum angles for the YZ plane. For shorter wavelengths, the overall two-fold symmetry was also maintained for the cleavage plane (XY) but became more isotropic for some peaks, as shown in Fig. S6. In particular, the intensity of $A_8$ obtained with 514 nm exhibited the least dependence on the azimuth with the min/max intensity ratio of 0.75. This mode is useful in quantifying the material because of its virtual isotropy, as shown in a recent photochemical study of $CrPS_4$.[26] On the other hand, the azimuth-dependence of Raman signals can be used in determining the crystallographic orientation of $CrPS_4$. Whereas full-angle polar graphs like Fig. 6 provide the best accuracy, the ratiometric analysis of a single spectrum will be satisfactory for general purposes. The calibration curves for the three wavelengths were generated for the XY plane and presented in Fig. S7, where the intensity ratio of $A_6/A_2$ (633 nm) and $A_4/A_6$ (514 and 457 nm) exhibited sufficient variation for practical applications.

**Complex Raman tensors of $CrPS_4$.** The distinctive angle-dependence of the Raman spectra reflects the anisotropic nature of $CrPS_4$, which can be quantitatively described by Raman tensors given in Table 1. Because $CrPS_4$ has significant light absorption at the excitation photon energies, however, the Raman tensors must be expressed in complex form as follows[45, 47]:

$$R(A)= \begin{pmatrix} be^{i\phi_b} & 0 & de^{i\phi_d} \\ 0 & ce^{i\phi_c} & 0 \\ de^{i\phi_d} & 0 & ae^{i\phi_a} \end{pmatrix} \quad (2)$$

$$R(B)= \begin{pmatrix} 0 & fe^{i\phi_f} & 0 \\ fe^{i\phi_f} & 0 & ee^{i\phi_e} \\ 0 & ee^{i\phi_e} & 0 \end{pmatrix} \quad (3)$$

, where $\phi_a \sim \phi_f$ represent the phase information of the tensor components introduced in Table 1. Then, the azimuth-dependence of the polarized Raman signals in Fig. 6 can be described using Equation 1. For the measurements of the XY plane in Fig. 1b, for example, the incident and scattered lights in the parallel configuration can be represented by

$$\hat{e}_i = (sin\theta, cos\theta, 0) \quad (4)$$

$$\hat{e}_s = (sin\theta, cos\theta, 0) \quad (5)$$

For the cross configuration, the angle in Equation 5 needs to be replaced by θ + 90°. By substituting these two vectors and the complex Raman tensors into Equation 1, the following four expressions can be obtained respectively for the Raman intensities in the parallel (∥) and cross (⊥) configuration for both types of Raman peaks:

$$I_A^{\parallel} \propto b^2 sin^4\theta + c^2 cos^4\theta + 2bc\,sin^2\theta cos^2\theta \cos\phi_{bc} \quad (6)$$

$$I_A^{\perp} \propto sin^2\theta cos^2\theta (b^2 + c^2 - 2bc \cos\phi_{bc}) \quad (7)$$

$$I_B^{\parallel} \propto [f\,sin(2\theta)]^2 \quad (8)$$

$$I_B^{\perp} \propto [f\,cos(2\theta)]^2 \quad (9)$$

, where $\phi_{bc}$ is the difference between $\phi_b$ and $\phi_c$. The angular dependence for the other two facets could also be derived similarly and was given in Supporting Information (Section C). As shown by the solid lines in Fig. 6 & Fig. S5, Equations 6 ~ 9 described very well all the experimental data obtained in both polarization configurations, which allowed determination of the Raman tensor values. Fitting the data in Fig. 6a using Equation 6, for instance, resulted in the ratio b/c and phase difference $\phi_{bc}$. In Table 2, we summarized b/c ratios and $\phi_{bc}$ of several A-type modes determined for the three excitation wavelengths. It is notable that $\phi_{bc}$ deviates significantly from zero for all modes except $A_6$ at 457 nm, which stresses the complex nature of the Raman tensors. Indeed, the polar graph for $A_7$ for 633 nm could not be fitted with real-valued Raman tensors (Fig. S8). It is also clear that the isotropic nature of the parallel signal of $A_8$ for 514 nm is associated with the fact that b/c is nearly unity and $\phi_{bc}$ is close to zero, judging from Equation 6.

The Raman tensors for excitation at 633 nm could be further evaluated with the data for the two edge planes (Fig. 6c & 6d). The XZ and YZ planes enabled evaluation of b/d and a/c ratios for A modes, respectively. The phase differences of $\phi_{bd}$ and $\phi_{ac}$ were also retrieved from the former and latter planes, respectively. The complete Raman tensors for several major A-type modes are given in Table 3 after normalizing each tensor with its component, $ae^{i\phi a}$. However, the two nonzero components of B modes (**e** and **f**) could not be determined because each is proportional to the overall Raman intensities of associated polarized Raman signal, as shown in Equations 8 & 9 and Supporting Information Section C.

As mentioned in the introduction, interest in $CrPS_4$ had been dormant until the recent research activities on various 2D materials. The lattice vibrations of bulk and few-layer $CrPS_4$ were first reported in 2017 by our group and collaborators,[22] but without systematical assignment by symmetry. All the peaks observed by J. H. Lee et al.[22] were confirmed and could be associated with one of 10 A or 6 B modes as summarized in Table S2. The current work also revealed 11 more peaks, including those at 85.1, 115.2, and 178.2 cm$^{-1}$. The Raman spectra of 1L $CrPS_4$ could be obtained only after careful passivation because of its ambient and photoinduced instability.[26] We also note that there were recent polarized Raman studies on bulk $CrPS_4$ by H. Wu et al.[38] and P. Gu et al.[36] Because our approach was not limited to the cleavage plane but extended to the two orthogonal edge planes, all possible components of Raman tensors could be determined. Moreover, the phase information of the tensors could also be retrieved by using complex tensor forms, which was found to be essential in describing the polarized Raman behavior correctly. Finally, the dependence of Raman signals on the excitation wavelength suggested strong electron-phonon coupling at higher photon energies.

**Conclusions**

We performed a polarized Raman study of layered magnetic semiconductor $CrPS_4$ using three different excitation wavelengths of 457, 514, and 633 nm. To obtain maximal information on its Raman tensors, we prepared three orthogonal crystallographic facets, including two edge planes. Among 33 Raman active modes (16 A and 17 B), 27 observed Raman peaks were assigned to 19 A and 8 B-type modes according to the polarization selection rule. The intensity of each Raman peak exhibited significant dependence on the angle between the light polarization and crystallographic axes. Moreover, the angle-dependent intensity patterns varied for different excitation wavelengths and crystallographic planes. Despite its apparent complexity, however, the polarization dependence of the Raman signals could be well described with complex-valued Raman tensors, which varied significantly for the Raman modes and excitation wavelengths. The ratiometric intensity analysis derived from the angle-resolved Raman data also allowed determination of crystallographic orientation with a single Raman spectrum.

**Methods**

**Sample preparations.** $CrPS_4$ crystals were grown by chemical vapor transport at 700 °C, as reported elsewhere.[22] Samples terminated with cleavage planes were prepared by mechanically exfoliating crystalline $CrPS_4$ flakes onto Si substrates with the 285-nm-thick thermally grown $SiO_2$ layers. Their typical thickness and the roughness ranges were 300 ~ 500 nm and 0.12 ~ 0.16 nm, respectively. Samples with well-defined edge planes orthogonal to the cleavage plane were prepared by mechanical cutting and polishing, as detailed in Fig. S2.

**Polarized Raman measurements.** All Raman spectra were obtained in a back-scattering geometry using a home-built micro Raman setup (Fig. S9) with three different excitation wavelengths of 633 (1.96), 514 (2.41), and 457 nm (2.71 eV) as explained elsewhere.[48] Briefly, the excitation laser beams of linear polarization were focused by an objective lens (40X, numerical aperture = 0.60) onto samples within a diffraction-limited spot of ~1.0 μm in diameter. Scattered signals were collected by the objective and fed to a spectrometer with a liquid nitrogen-cooled charge-coupled detector. For polarized measurements in the parallel and cross configurations, Raman signals of a specific linear polarization were selected with an analyzing polarizer and converted into circularly polarized beams with a broadband quarter-wave plate to avoid the polarization dependence of the detector unit. To vary the azimuthal angle, we rotated samples in a step of 10° using a rotational mount. The overall spectral accuracy was 0.5 cm$^{-1}$, and the spectral resolution defined by FWHM of the Rayleigh peak was 3 cm$^{-1}$. We maintained the average power of the excitation beams below 2 mW to avoid photoinduced damage in samples.

**ASSOCIATED CONTENT**

**Supporting Information.** This material is available free of charge at http://pubs.acs.org.

Topography of three orthogonal crystallographic facets, preparation of edge planes by sawing and polishing process, resolution of $A_1$ and $B_2$ peaks using a grating with a higher groove density, full-angle polarized Raman spectra obtained from XY-planed $CrPS_4$, angular dependence of B-mode Raman intensities of XY-planed $CrPS_4$, wavelength-dependence of intensity polar graphs of XY-planed $CrPS_4$, polar graphs of selected intensity ratios for determination of crystallographic orientation, real vs complex Raman tensors, scheme of polarized Raman spectroscopy setup; polarization selection rule of $C_2$ point group for back-scattering geometry, assignment of Raman modes compared with the first Raman study; supporting equations.

**AUTHOR INFORMATION**

**Corresponding Author**

*E-mail: sunryu@postech.ac.kr

**Author Contributions**

The manuscript was written through the contributions of all authors. All authors have given approval to the final version of the manuscript.

**Notes**

The authors declare no conflict of interest.

**ACKNOWLEDGMENTS**

We thank Bheema Lingam Chittari, Jeil Jung, Onnuri Kim, Moon Jeong Park, Jungcheol Kim, and Hyeonsik Cheong for productive discussion of this research. This work was supported by the National Research Foundation of Korea (NRF- 2020R1A2C2004865, 2019R1A4A1027934, 2020R1A2C2014687).

**Figures and captions**

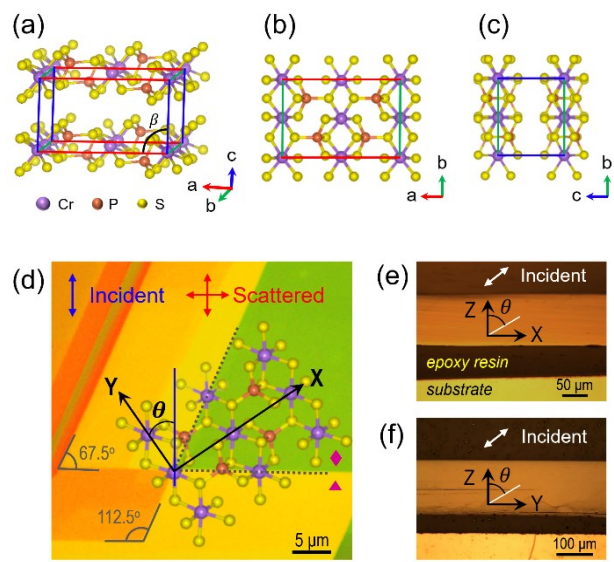

**Figure 1. Crystalline structure and three orthogonal facets of CrPS$_4$.** (a) Monoclinic unit cell of CrPS$_4$ spanned with three crystallographic axes of *a*, *b,* and *c* in the schematic structural model. (b ~ d) Optical micrographs of XY (a), XZ (b), and YZ (c) planes of bulk CrPS$_4$, respectively. X, Y, and Z axes denote *a*, *b*, and a surface normal to *ab* plane, respectively. XY planes generated by mechanical cleaving contain parallelograms with preferential angles as shown in (d). X axis is parallel to the bisector of the acute angle.[22] The azimuth (θ) was defined as an angle between the polarization direction of the incident beam and the Y or Z axis. The thickness of the green (denoted with a diamond) and yellow (denoted with a triangle) regions were about 120 and 70 nm, respectively.

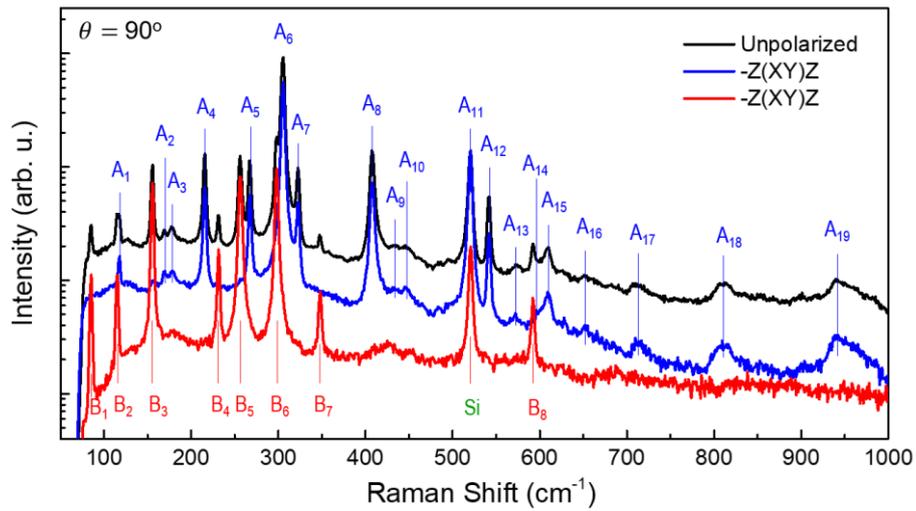

**Figure 2. Symmetry assignment of Raman peaks of CrPS$_4$.** All Raman peaks of the unpolarized spectrum (black) could be detected in the parallel (blue) or cross (red) polarization configurations in a mutually exclusive manner. The Raman peak denoted Si in the cross spectrum originated from underlying Si substrate. Optical configurations are specified in the legend using the Porto's notation: The first (last) letter designates the propagation direction of the incident (scattered) light, and the second (third) letter the direction of polarization of the incident (scattered) light (see Table S1 for polarization selection rule). The spectra were obtained with 633 nm excitation wavelength at θ = 0°.

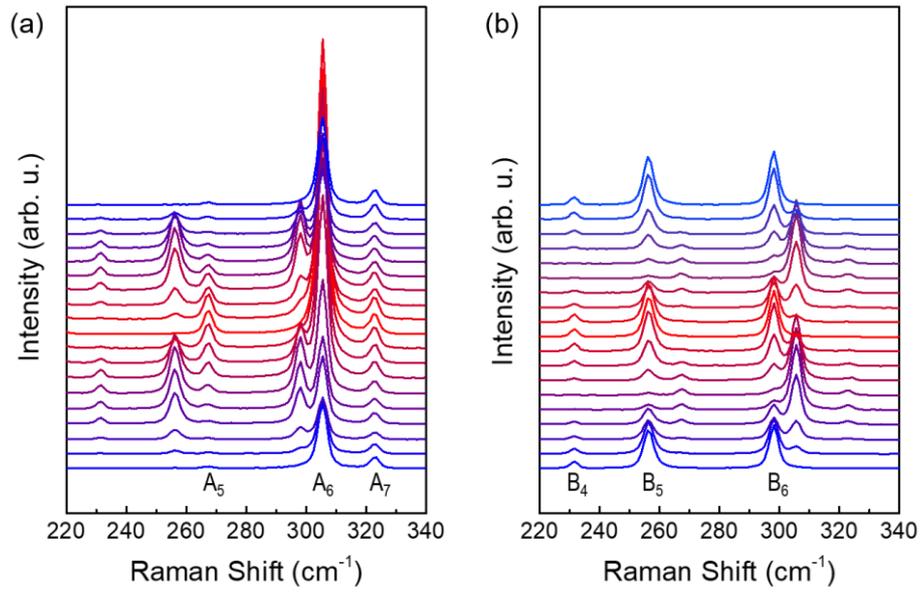

**Figure 3. Full-angle polarized Raman spectra of XY-planed CrPS$_4$.** (a & b) Spectra obtained with 633 nm in parallel (a) and cross (b) configurations for varying θ. Samples were rotated about Z axis in step of 10° from 0 to 180° (see Fig. S4 for the full-range spectra obtained for 0 ~ 360°).

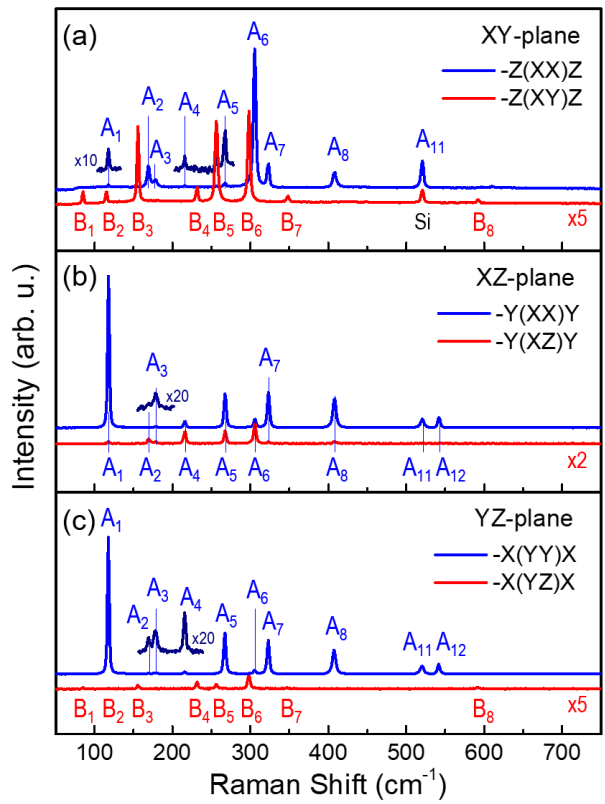

**Figure 4. Polarized Raman spectra of three orthogonally faceted crystals.** (a ~ c) Polarized Raman spectra obtained from XY (a), XZ (b), and YZ (c) facets of CrPS$_4$. The data were obtained with 633 nm excitation wavelength at θ = 0°

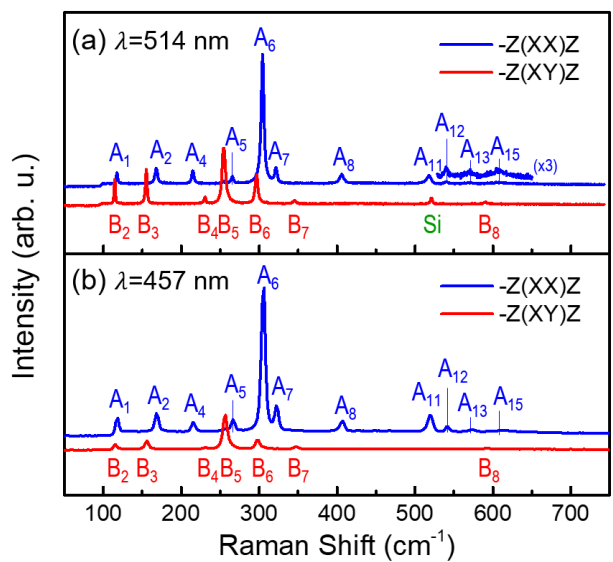

**Figure 5. Dependence of intensity profiles on excitation wavelength.** (a & b) Polarized Raman spectra obtained with 514 (a) and 457 nm (b) for XY facet of $CrPS_4$ ($\theta = 0°$).

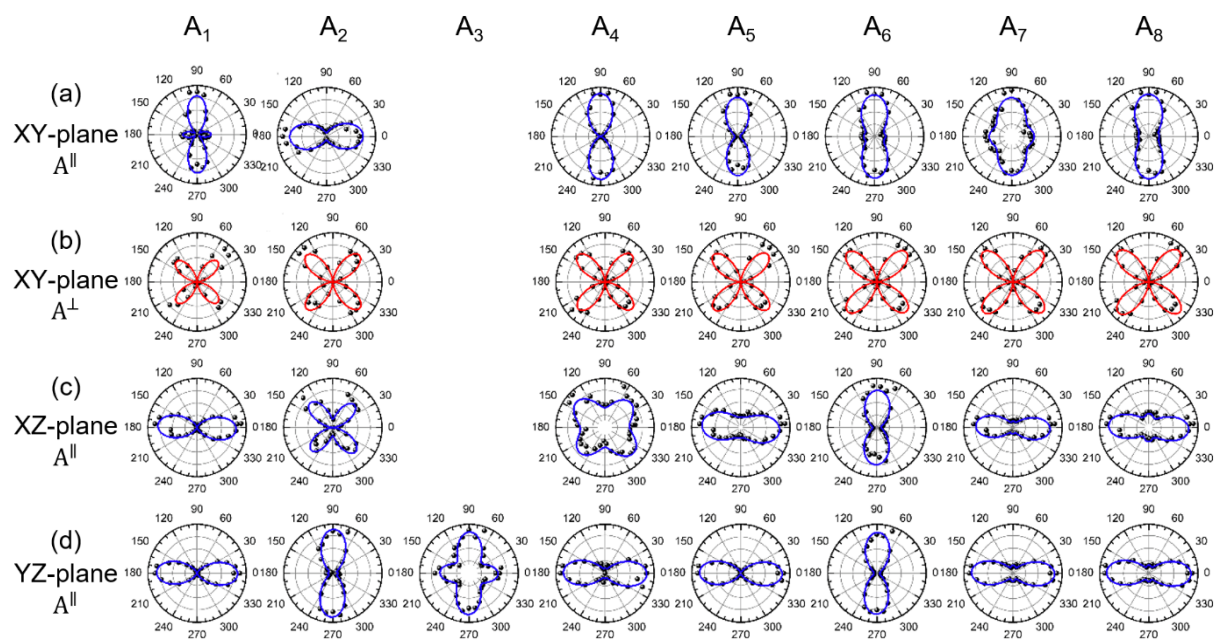

**Figure 6. Complex Raman tensors describing intricate polarization dependence.** (a & b) Polar graphs of parallel (a) and cross (b) Raman signals of major A-symmetry peaks from XY plane (see Fig. S5 for B-symmetry peaks). (c & d) Polar intensity graphs of A-type peaks obtained in parallel configuration from XZ (c) and YZ (d) planes. All data were obtained with 633 nm (see Fig. S6 for other wavelengths). Solid lines are fits based on complex Raman tensors (Equations 2 ~ 3). See Equations 6 ~ 9 & Supporting Equations for the mathematical functions used for the fitting.

**Tables**

**Table1. Generic Raman tensors of CrPS$_4$ (C$_2$ point group)**

| Mode | A | B |
|---|---|---|
| Raman tensor | $\begin{pmatrix} \tilde{b} & 0 & \tilde{d} \\ 0 & \tilde{c} & 0 \\ \tilde{d} & 0 & \tilde{a} \end{pmatrix}$ | $\begin{pmatrix} 0 & \tilde{f} & 0 \\ \tilde{f} & 0 & \tilde{e} \\ 0 & \tilde{e} & 0 \end{pmatrix}$ |

*The tensor components are complex-valued in general

**Table 2. Tensor components retrieved from XY-plane data**

| | 457 nm | | | 514 nm | | | 633 nm | | |
|---|---|---|---|---|---|---|---|---|---|
| | b/c | phase (°) | | b/c | phase (°) | | b/c | phase (°) | |
| A$_1$ | - | - | - | 1.14 | 55 | ± 6.3 | 1.66 | 139 | ± 7.3 |
| A$_2$ | - | - | - | 0.26 | 81 | ± 9.4 | 0.32 | 10 | ± 108.2 |
| A$_4$ | 1.94 | 29 | ± 12.4 | 1.40 | 22 | ± 15.6 | 7.44 | 94 | ± 12.4 |
| A$_5$ | 1.77 | 24 | ± 11.8 | - | - | - | 3.56 | 84 | ± 13.1 |
| A$_6$ | 0.52 | 0 | - | 0.57 | 39 | ± 10.0 | 1.95 | 77 | ± 5.4 |
| A$_7$ | 0.93 | 57 | ± 3.8 | 0.81 | 10.62 | ± 38.4 | 1.31 | 67 | ± 5.9 |
| A$_8$ | 1.21 | 22 | ± 12.1 | 0.99 | 19.1 | ± 19.4 | 2.02 | 66 | ± 6.7 |

**Table 3. Complete Raman tensors of major A modes**

| | R(A) | | R(A) |
|---|---|---|---|
| $A_1$ | $\begin{pmatrix} 12.50e^{1.11\pi i} & 0 & 2.50e^{1.69\pi i} \\ 0 & 9.09e^{0.62\pi i} & 0 \\ 2.50e^{1.69\pi i} & 0 & 1 \end{pmatrix}$ | $A_6$ | $\begin{pmatrix} 0.34e^{1.76\pi i} & 0 & 0.07e^{0.41\pi i} \\ 0 & 0.28e^{0.46\pi i} & 0 \\ 0.07e^{0.41\pi i} & 0 & 1 \end{pmatrix}$ |
| $A_2$ | $\begin{pmatrix} 0.03e^{0.49\pi i} & 0 & 1.89e^{1.69\pi i} \\ 0 & 0.34e^{0.54\pi i} & 0 \\ 1.89e^{1.69\pi i} & 0 & 1 \end{pmatrix}$ | $A_7$ | $\begin{pmatrix} 2.38e^{0.48\pi i} & 0 & 0.12e^{1.66\pi i} \\ 0 & 2.27e^{0.54\pi i} & 0 \\ 0.12e^{1.66\pi i} & 0 & 1 \end{pmatrix}$ |
| $A_4$ | $\begin{pmatrix} 1.11e^{0.03\pi i} & 0 & 0.85e^{1.49\pi i} \\ 0 & 2.86e^{0.46\pi i} & 0 \\ 0.85e^{1.49\pi i} & 0 & 1 \end{pmatrix}$ | $A_8$ | $\begin{pmatrix} 1.67e^{0.48\pi i} & 0 & 0.09e^{1.58\pi i} \\ 0 & 2.27e^{0.50\pi i} & 0 \\ 0.09e^{1.58\pi i} & 0 & 1 \end{pmatrix}$ |
| $A_5$ | $\begin{pmatrix} 1.92e^{0.29\pi i} & 0 & 0.14e^{1.51\pi i} \\ 0 & 5.56e^{0.56\pi i} & 0 \\ 0.14e^{1.51\pi i} & 0 & 1 \end{pmatrix}$ | | |

**TOC Figure**

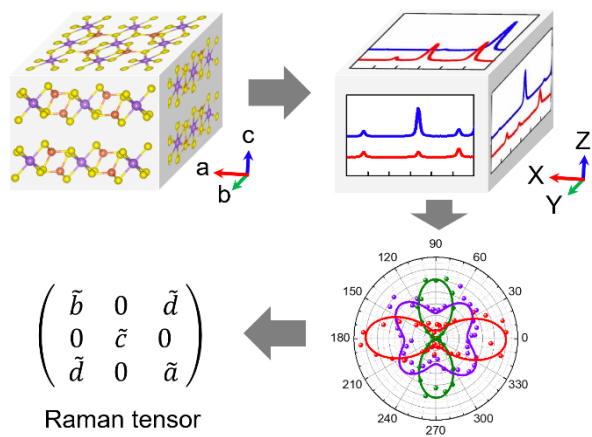

Raman tensor